\documentstyle[amssymb,preprint,aps]{revtex}

\newtheorem{Theorem}{Theorem}
\newtheorem{Lemma}[Theorem]{Lemma}
\newcommand{\Proof}{\noindent {\em Proof:}\ \ }
\newcommand{\QED}{\hfill QED.\medskip\par}

\newcommand{\bR}{{\Bbb R}}
\newcommand{\bC}{{\Bbb C}}
\newcommand{\frt}{{\textstyle{\frac{1}{\sqrt{2}}}}}
\newcommand{\half}{{\textstyle{\frac{1}{2}}}}
\newcommand{\quarter}{{\textstyle{\frac{1}{4}}}}
\newcommand{\imu}{{\rm i}\,}
\newcommand{\edth}{\eth}
\newcommand{\edthbar}{\bar{\eth}}
\newcommand{\mbar}{\bar{m}}
\newcommand{\cD}{{\cal D}}
\newcommand{\cN}{{\cal N}}
\newcommand{\cT}{{\cal T}}
\newcommand{\cL}{{\cal L}}
\newcommand{\Qp}{Q^{+}}
\newcommand{\Qm}{Q^{-}}
\newcommand{\Qpbar}{\bar{Q}^{+}}
\newcommand{\Qmbar}{\bar{Q}^{-}}
\renewcommand{\div}{{\mbox{div}}}
\newcommand{\curl}{{\mbox{curl}}}
\renewcommand{\Re}{{\mbox{Re}}}

\begin{document}
\draft
%
\preprint{UNE-MSCS-96-127, gr-qc/9611045}
\title{Einstein Equations in the Null Quasi-spherical Gauge}
\author{Robert Bartnik}
\address{
School of Mathematics and Statistics\\
University of Canberra\\
Belconnen, ACT 2616, Australia
}
\date{\today}
\maketitle
\begin{abstract}
Properties of the Einstein equations in a coordinate gauge based on 
expanding null hypersurfaces foliated by metric 2-spheres are described.
This null quasi-spherical (NQS) gauge leads to particularly simple analyses 
of the characteristic structure of the equations and of the propagation of 
gravitational shocks, and clarifies the geometry of timelike boundary 
condition.
A feature of the NQS gauge is
the use of the standard $\eth$ (``edth'') operator on $S^{2}$ 
to express angular derivatives, and the consequent 
use of spin-weighted spherical harmonic decompositions of the metric fields.
\end{abstract}


The objectives of numerical relativity in particular, force us to address 
the problem of constructing \emph{explicit} formulations of the full 
Einstein equations.  There are two main avenues of approach, respectively 
using either a 3+1 ADM representation \cite{ADM62}, which leads to a 
Cauchy initial value problem, or a characteristic coordinate, an approach 
first considered by Bondi \cite{Bondi62,Sachs61}.

Both the second order and the recently popular first order 
\cite{BMES96,AACY95,FR96} formulations of the 3+1 equations involve 
coordinate gauge freedoms which must be eliminated through gauge-fixing 
conditions which typically involve subsidiary differential equations.  
First order formulations in particular introduce large numbers of auxiliary 
variables to parameterise the field, and these variables must satisfy 
additional compatibility relations.  Whilst these features do not present 
difficulties of a conceptual or theoretical nature, they do cause 
difficulties in the practical numerical formulation of the equations --- in 
disentangling geometric from coordinate gauge and numerical algorithm 
effects \cite{BMES96}, in formulating the numerical algorithm to ensure 
that the constraints and compatibility relations are preserved by the 
evolution \cite{Evans88,Gundlach96}, and in simply constructing  codes 
handling large number of variables.

Characteristic methods \cite{Sachs62},\cite{NU62} on the other hand 
typically involve very little coordinate freedom and lead to formulations 
of the equations involving a small number of variables, with few 
compatibility conditions.  However, the advantage of simpler 
representations of the equations is balanced by the disadvantage that 
characteristic coordinates are generically not globally applicable, for 
either topological or caustic reasons.  However, a characteristic-based 
coordinate system can be very useful when available --- for example, in 
describing asymptotic structure \cite{Bondi62}, or for describing 
discontinuity properties of vacuum spacetimes \cite{Stellmacher38}.

The Newman-Penrose formulation in terms of null tetrad based connection 
(``spin'') coefficients \cite{NP62} can be used to describe the 
characteristic structure of the Einstein equations, either by regarding the 
equations as determining the propagation of the spin coefficients 
\cite{Papapetrou71,Edgar89}, or by introducing coordinates and determining 
the spin coefficients in terms of the metric parameters (eg. \cite{NU62}).
The former (``pure NP'') technique introduces numerous compatibility 
differential relationships amongst the spin coefficients, which suggests 
that more variables have been introduced than are necessary to 
fully describe the connection.  
The latter approach bridges the gap between coordinate- and 
connection-based techniques, and is the starting point for this paper.  

Ideally, a ``good'' parameterisation of the Einstein equations will have 
limited or no gauge freedom (to suppress unphysical gauge modes in the 
numerical evolution, for example \cite{BMES96}), with parameters free of 
constraints and having a direct relation to known radiation parameters, and 
leading to simple expressions for the Einstein equations, to facilitate 
analysis of the local existence, decay etc.

The null quasi-spherical gauge presented here is a modification of the 
usual characteristic formulations \cite{Bondi62},\cite{NU62} which 
satisfies many of these criteria.  It combines a characteristic gauge 
(exploiting the ode structure of the Einstein on a null hypersurface 
\cite{Sachs62}) with the reduced gauge-dependence of the null tetrad 
methods \cite{NP62}, but avoids the complications of the spin coefficient 
integrability conditions.  The quasi-spherical condition means that all 
derivatives can be presented using radial and time partial derivatives, 
combined with canonical differential operators on $S^{2}$ with the standard 
metric.

An interesting feature is the central role played by the NQS metric field 
$\beta$, which determines in particular both the shear $\sigma_{NP}$ of the 
outgoing null curves of the null coordinate, and the intrinsic geometry of 
the null hypersurfaces.  Given $\beta$ on one null hypersurface, the 
remaining NQS parameters are determined by integrating a system of radial 
ordinary differential equations which generalise the usual null 
Raychaudhuri equation (which is just the first equation (\ref{Gll:eq}) of 
the system).  Heuristically, $\beta$ may be regarded as describing the 
ingoing gravitational radiation passing through the outgoing null hypersurfaces.

The radial system of equations for the metric parameters is of course 
well-known in similar situations \cite{Bondi62,NU62} and can be 
understood in terms of the NP formulation of the equations for Einstein 
tensor terms $\Phi_{00}, \Phi_{01}, \Phi_{02}$ and $\Phi_{11}+3\Lambda$, 
which can be expressed solely in terms of derivatives tangent to the null 
hypersurfaces of the NP spin coefficients.

However, closer examination of the explicit NQS coordinate representations 
shows that the hypersurface equations may be expressed in terms of 
hypersurface-tangential derivatives of the metric functions and just a {\em 
subset} of the connection parameters (see 
Eqns.~(\ref{Gll:eq}--\ref{Gmm:eq}) below).  In all, we obtain a complete 
parameterisation of both the metric and the connection in terms of just 
$12$ real variables, made up of 2 real and 2 complex variables for each of 
the metric ($u,v,\beta,\gamma$) and the connection ($H,J,Q,K$).  Not only 
is this cheaper than the NP parameterisation (which requires 12 complex 
variables for the connection alone), but it is also {\em explicit} in the 
sense that there are no auxiliary compatibility conditions.  The resulting 
expressions for the Einstein tensor (\ref{Gll:eq}--\ref{Gmm:eq}), 
(\ref{eq:Gnn}--\ref{eq:Gnm}), (\ref{eq:Gmb}) and Weyl tensor (\ref{Psi0}--\ref{Psi4}) 
components are comparable in simplicity with the usual NP expressions, yet 
have the advantage that they are completely explicit and involve fewer 
variables.

The reduced set $(H,J,K,Q)$ of connection parameters was found by careful 
examination of the explicit expressions for the hypersurface Einstein 
tensor components in the NQS metric.  It seems quite likely that a similar 
examination using instead other characteristic-based metric forms 
(eg.~Bondi, Newman-Unti) will uncover other reduced (``beyond 
NP'') parameterisations for the connection.

Because of the very explicit form of the Einstein equations in the NQS
gauge, it enables us to give simple descriptions of several relatively
well-known computations, such as: the null hypersurface matching
conditions and the propagation laws for gravitational (vacuum) shock
waves \cite{Stellmacher38,Penrose72}; a description of the free
boundary data for the Einstein equations on a timelike boundary
surface; and the relationship between the outgoing Einstein tensor
component and the geodesic parameterisation of the outgoing null rays.
Two applications which are not discussed here, but which are quite
feasible within the NQS framework, are the rederivation of the
linearised Einstein equations \cite{RW57,Zerilli70,Spillane94}, and
the asymptotic form of a vacuum NQS metric.  The application which
motivated the present study, namely the numerical evolution of of a
vacuum black hole spacetime in the NQS gauge, will be described
elsewhere \cite{RAB96a}, \cite{Norton96}.

\section{NQS Einstein equations}
The central idea is to assume that the radial function foliates the null 
hypersurfaces by {\em metric} spheres, of area $4\pi r^{2}$.  
A similar foliation was exploited in \cite{RAB93} to construct spatial initial 
data satisfying the Hamiltonian constraint.
Denoting the null coordinate by $z$ (cf.~\cite{Stellmacher38}) 
and letting $\vartheta,\varphi$ denote standard 
polar coordinates on the spheres, the most general compatible metric (assuming 
that $r$ is a valid coordinate) is
\footnote{Our metric has signature $+2$, the expressions below for the 
NP spin coefficients have been adjusted to give the usual signs, and the 
curvature tensor sign convention has been chosen so that 
$G_{\ell\ell}=2\Phi_{00}$ is non-negative for physically reasonable matter.}
\begin{equation}
        ds^{2}_{NQS} = -2u\,dz(dr+v\,dz) +
        2\,|r\bar{\theta}+\beta\,dr+\gamma\,dz|^{2},
    \label{ds2:nqs}
\end{equation}
where we use a complex notation for simplicity, with
\[    \theta  =  \frt(d\vartheta + \imu\sin\vartheta\, d\varphi),
\]
\vspace{-20pt}
\[
    \beta = \frt (\beta^{1}-\imu\beta^{2}),\qquad
    \gamma = \frt (\gamma^{1}-\imu\gamma^{2}).
\]
Thus $u,v$ and $\beta,\gamma$ are metric parameters, with $u,v$ real and 
$u>0$, and we may regard $\beta,\gamma$ either as spin-1 (complex) fields 
or as vector fields tangent to the spheres, via
$    \beta \sim   \beta^1 \partial_{\vartheta} + 
   \beta^2 \csc\vartheta\,\partial_{\varphi} $.

The metric form Eqn.~(\ref{ds2:nqs}) differs from other parameterisations 
of the metric with a characteristic coordinate 
\cite{Sachs61},\cite{Bondi62}, \cite{NP62}, \cite{NU62},\cite{Hawking68} 
principally in the choice of radial coordinate.  For example, Bondi-Sachs 
use a luminosity parameter determined by the volume form in the angular 
directions \cite{CMS91} (with the consequent disadvantage that the 
luminosity coordinate depends strongly on the {\em specific} choice of 
angular coordinate labelling of the outgoing null curves), and Newman-Unti 
use the geodesic affine parameter (and thus the radial coordinate is not 
determined uniquely from the intrinsic induced metric on the null 
hypersurfaces).  However, the NQS radius and radial curves are determined 
by the metric $S^{2}$ foliation and standard polar coordinates, so that the 
NQS radial curves $(z,\vartheta,\varphi) = const.$ do not in general 
coincide with the null generating curves.

For the purposes of this paper it is sufficient to regard the metric form 
(\ref{ds2:nqs}) as an ansatz (assumption) rather than as a coordinate 
condition.  The important question of whether this ansatz is {\em 
generic}, in the sense that coordinates satisfying the NQS conditions may 
be found in all metrics in some open subset of the space of ``all'' 
metrics, does not yet have a complete answer.  Some of the evidence which 
suggests that the NQS coordinates are generically available is discussed 
below; we note also that the gauge used by Regge, Wheeler and Zerilli 
\cite{RW57,Zerilli70} in the 
study of the linearisation of the Einstein equations about Schwarzschild, 
is just a linearised NQS gauge.  Thus the NQS gauge could be used to extend
the Regge-Wheeler-Zerilli perturbation analysis to include nonlinear effects.
(The RWZ equations in the NQS gauge were rederived in \cite{Spillane94}.)

The complex notation suggests encoding the angular derivatives using
the differential operator $\eth$ 
(``edth'' \cite{PR84},\cite{ET82}), defined on a spin-$s$ field $\phi$ by
\begin{equation}
    \eth \phi = \frt \sin^{s}\!\vartheta\left(
             \frac{\partial }{\partial \vartheta} 
           - \frac{\imu}{\sin\vartheta} \frac{\partial }{\partial \varphi}
      \right)\left(\sin^{-s}\!\vartheta\,\phi\right).
\end{equation}

All the standard differential operators on $S^{2}$ may be expressed in 
terms of $\eth$, for example:
\begin{eqnarray*}
    \div\beta & = & \eth\bar{\beta} + \bar{\eth}\beta = 
    \beta^{1}_{;1}+\beta^{2}_{;2}
\\
    \curl\beta & = & \imu(\bar{\eth}\beta-\eth\bar{\beta})
        = \beta_{1;2}-\beta_{2;1}
\\
    \Delta \phi & = & (\eth\bar{\eth}+\bar{\eth}\eth)\phi.
\end{eqnarray*}
Here the indices refer to the $S^{2}$-orthonormal frame
$\partial_{\vartheta}$, $\csc\vartheta\,\partial_\varphi$, 
and the standard $S^{2}$ covariant derivative.  The spin-weighted 
spherical harmonics $Y_{slm}$ are then eigenfields of the Laplacian, 
satisfying $\Delta Y_{slm}= (s^{2}-l(l+1))Y_{slm}$, 
$|s|\le l$, and $Y_{slm}$ may be taken proportional to 
$\eth^{s}Y_{lm}$ for $s\ge0$ and $\bar{\eth}^{-s}Y_{lm}$ for $s<0$, 
where $Y_{lm}$ are the usual spherical harmonic functions.

To compute the curvature of $ds^{2}_{NQS}$ we introduce the vector frame
$(\ell,n,m,\mbar)$, null with respect to $ds^{2}_{NQS}$:
\begin{eqnarray}
        \ell &=& \partial_{r}-r^{-1}\partial_{\beta},
\nonumber\\
        n &=& u^{-1}(\partial_{z}-r^{-1}\partial_{\gamma}
                -v(\partial_{r}-r^{-1}\partial_{\beta})),
\label{tetrad:def}
\\
        m&=&\frac{1}{r\sqrt{2}}
        (\partial_{\vartheta}-\imu\csc\vartheta\,\partial_{\varphi}),
\nonumber
\end{eqnarray}
and its associated dual (null) coframe
$dr+v\,dz$, $u\,dz$, $r\,\theta+ \bar{\beta}\,dr +\bar{\gamma}\,dz$ and
$r\,\bar{\theta}+ {\beta}\,dr +{\gamma}\,dz$.
Note that $\partial_{\beta}$ denotes the angular directional derivative
$\beta^{1}\partial_{\vartheta}+\beta^{2}\csc\vartheta\,\partial_{\varphi}$,
which coincides with the $S^{2}$-covariant derivative 
$\nabla_{\beta}=\bar{\beta}\eth+\beta\bar{\eth}$ when acting on functions.  

It will be useful to introduce the differential operators
\begin{eqnarray*}
  \cD_{r} &:=& \partial_{r} - r^{-1}\nabla_{\beta} = \partial_{r} - 
  r^{-1}(\beta\bar{\eth} + \bar{\beta}\eth),\\
  \cD_{z} &:=& \partial_{z} - r^{-1}\nabla_{\gamma} =\partial_{z} - 
  r^{-1}(\gamma\bar{\eth} + \bar{\gamma}\eth),
\end{eqnarray*}
which have a natural interpretation as spin weight 0 operators.  Since we 
consider here only integer spins, we may use the complexified cotangent 
bundle $T^{*(1,0)}_{\bC}S^{2}$ as a model for the spin-1 line bundle 
$\cL^{1}$.  
{}From $\cL^{1}$ we construct the spin-$s$ complex line bundle $\cL^{s}$ 
over $\bR\times\bR^{+}\times S^{2}$, 
whose sections are just spin-$s$ fields depending also 
on the parameters $(z,r)$. 
Then the action of $\cD_{r}$, $\cD_{z}$ on spin-$s$ fields 
is defined via the standard 
covariant derivative on $S^{2}$, extended naturally to 
$T^{*(1,0)}_{\bC}S^{2}$ and $\cL^{s}$.

The structure of the hypersurface equations (\ref{Gll:eq}-\ref{Gmm:eq}) 
motivates the introduction of the NQS connection parameters
$H,J,K,Q,Q^{\pm}$, defined by:
\begin{eqnarray}
H  &=& u^{-1}(2- \div\beta), \label{def:H} \nonumber \\
J  &=& \div \gamma + v(2-\div\beta) , \label{def:J}\nonumber \\
K  &=& v\edth \beta - \edth \gamma, \label{def:HJKQ}\\
Q  &=& r\cD_{z}\beta-r\cD_{r}\gamma +\gamma, \label{def:Q}\nonumber\\
Q^{\pm} &=& u^{-1}(Q \pm \edth u). \label{def:Qpm}\nonumber 
\end{eqnarray}
These have well-defined spin-weights under $S^{2}$ frame 
rotations: $u,v,H,J$ have spin 0 (and are real), $\beta,\gamma,Q,Q^{\pm}$ 
have spin 1, and $K$ has spin 2.

The connection $\hat{\nabla}$ of $ds^{2}_{NQS}$ 
may be described in terms of the NQS parameters  
either via the Cartan connection coefficients 
$\omega_{abc}=g(a,\hat{\nabla}_{c}b)$, 
$a,b,c=\ell,n,m,\mbar$, or via the Newman-Penrose spin coefficients
\cite{NP62}:
\begin{eqnarray*}
\kappa_{NP} &=&  \omega_{\ell m\ell} = 0 
\\
\sigma_{NP} &=& \omega_{\ell mm}   = r^{-1}\edth \beta
\\
\rho_{NP}   &=& \omega_{\ell m\mbar} = {} -\half r^{-1}(2-\div\beta) 
\\
\epsilon_{NP} &=& \half(\omega_{\ell n\ell} + \omega_{\mbar m\ell})
   = \half u^{-1}\cD_{r}u 
{}+\imu r^{-1}\left(\frt\cot\vartheta\, \mbox{Im}(\beta) + 
                        \quarter\curl \beta  \right)  
\\
\gamma_{NP}&=& \half(\omega_{\ell nn} + \omega_{\mbar mn})
\\
  & =&
  \half u^{-1}\cD_{r}v
{}+\imu r^{-1}u^{-1}\left(\frt\cot\vartheta\, 
           \mbox{Im}(\gamma-v\beta) + 
                        \quarter(\curl\gamma-v\curl \beta)  \right)
\\
\bar{\alpha}_{NP} + \beta_{NP} &=& \bar{\pi}_{NP} = \omega_{\ell nm} 
   = \omega_{mn\ell} = \half r^{-1}\Qp  
\\
\bar{\alpha}_{NP} - \beta_{NP} &=& \omega_{m\mbar m} 
   = {}-\frt r^{-1}\cot\vartheta
\\
\tau_{NP} &=& \omega_{\ell mn}= \half r^{-1} \Qm
\\
\bar{\lambda}_{NP} &=& \omega_{mnm} = r^{-1}u^{-1}K
\\
\mu_{NP} &=&\omega_{\mbar nm} = {}-\half r^{-1}u^{-1}J
\\
\bar{\nu}_{NP}&=& \omega_{mnn} = r^{-1}u^{-1}\edth v
\end{eqnarray*}

As noted above, the NQS parameters $(u,v,\beta,\gamma)$ and $(H,J,K,Q)$ 
form a more efficient and complete representation of the metric and connection 
than the NP coefficients $\alpha_{NP},\ldots,\tau_{NP}$, in part because 
they are defined in terms of a specific 
coordinate-adapted frame $(\ell,n,m,\bar{m})$.  The ``compatibility'' 
conditions are also much simpler than those required of the NP 
coefficients.  Clearly, $H,J,K,Q$ are
 determined uniquely from the metric parameters $u,v,\beta,\gamma$.
Conversely, if $\beta,H,J,K$ are given then the remaining metric 
parameters $(u,v,\gamma)$ may be reconstructed as follows: 
$u$ is found algebraically from $\div\beta$ and $H$; 
$\gamma$  is found by solving
\begin{equation}
    \cL_{\beta}\gamma := 
    \eth \gamma + \frac{\eth \beta}{2 - \div \beta}\,\div \gamma 
    = J  \frac{\eth \beta}{2 - \div \beta} - K
\label{Dbeta:def}
\end{equation}
on each $S^{2}$;
and finally
$v= (J-\div\gamma)/(2-\div\beta)$.
Observe that the reconstruction process is local to the null 
hypersurfaces, since $\cD_{z}$ derivatives are not involved, and that
$Q$ is not used --- it 
is used instead to determine $\partial\beta/\partial z$.  

If $\beta=0$ then $\cL_{\beta}=\cL_{0}=\eth$, an elliptic operator on $S^{2}$ 
acting on spin-1 fields, which is surjective and has 6-dimensional
kernel consisting 
of the $l=1$ spin-1 spherical harmonics $Y_{11m}$, $m=-1,0,1$.
Consequently if $\beta$ is small (and we will show elsewhere that the 
pointwise  bound 
\begin{equation}\label{ebsize:ineq}
   |\eth\beta|<3^{-1/2}(2-\div\beta)
\end{equation}
is sufficient),
then (\ref{Dbeta:def}) is solvable for $\gamma$, uniquely if 
the $l=1$ components of $\gamma$ are prescribed.
The kernel of $\cL_{0}$ may be identified with the 
infinitesimal Lorentz transformations and conformal motions of $S^{2}$ --- 
this suggests we may interpret this ambiguity in the reconstruction of 
$\gamma$ as a gauge degeneracy.  

Observe that the hypersurfaces are expanding, $\rho_{NP}<0$, exactly when 
$\div\beta<2$, and this condition follows from (\ref{ebsize:ineq}).  The 
deformation term $\edth\beta/(2-\div\beta)$ in $\cL_{\beta}$ 
(\ref{Dbeta:def}) is equal to $-\half\sigma_{NP}/\rho_{NP}$, which is an 
invariant of the null geometry (type I in the terminology of 
\cite{Penrose72}).

To understand the genericity of the NQS ansatz, consider the effect of an 
infinitesimal variation $h_{ab}$ of the null metric $ds^{2}_{NQS}|_{\cN}$ 
on a null hypersurface $\cN$.   A calculation shows that the metric $S^{2}$ 
condition is preserved (at the linearised level) if the accompanying 
coordinate deformation vector $X=r(\zeta \mbar+\bar{\zeta}m)+rf\ell$ 
satisfies 
\begin{equation}
        \cL_{\beta}\zeta = {}-\half h_{0} \frac{\edth\beta}{2-\div\beta} 
                           -\half h_{2},
\label{Dbzeta}
\end{equation}
where $h_{0}=h_{ab}m^{a}\mbar^{b}$, $h_{2}=h_{ab}m^{a}m^{b}$, and 
\[
  f = {}-\frac{\div\zeta+\half h_{0}}{2-\div\beta}.
\]
Thus if the size condition (\ref{ebsize:ineq}) holds then the 
linearisation of the NQS gauge can be enforced by solving (\ref{Dbzeta}) for
 $\zeta$;
moreover, the existence of NQS coordinates in axially symmetric Bondi 
metrics was established in \cite{Spillane94}.  Consequently it is plausible to 
regard the NQS gauge as a coordinate condition rather than an ansatz.
Note also that the NQS parameters $u,v$, $\beta,\gamma$  
have the degrees of freedom ($6=10-4$)
expected of a general metric after removing all coordinate degeneracies.

The characteristic structure of the Einstein equations 
\cite{Sachs62} becomes particularly clear
when they are expressed in NQS coordinates using the connection quantities  
$H,J,K,Q$.  
Let $G_{ab}=R_{ab}-\half Rg_{ab}$ be the Einstein tensor, normalised by
$G_{\ell\ell}=\ell^{a}\ell^{b}G_{ab}=2\Phi_{00}$.  Then
$G_{\ell\ell}$, $G_{\ell m}$, $G_{\ell n}$, $G_{mm}$ are given by the
{\em hypersurface equations}  \cite{Sachs62}
\begin{eqnarray}
r \cD_{r}H&=&
   \left(\half\div\beta -
       \frac{2|\edth\beta|^2+r^{2}G_{\ell\ell}}{2-\div\beta} \right)H,
\label{Gll:eq}
\\
r\cD_{r}\Qm &=&
  (\edth\bar{\beta}-uH)\Qm + \Qmbar\edth\beta + 2\edthbar\edth\beta
{}+ u\edth H - H\edth u  + 2r^{2} G_{\ell m} ,
\label{Glm:eq} 
\\
r \cD_{r}J &=&
   -( 1-\div\beta)J + u - \half u |\Qp |^2
     {}-\half u\,\div(\Qp ) - u r^2 G_{\ell n} ,
\label{Gln:eq} 
\\
r \cD_{r}K &=&
\left(\half \div\beta + \imu\curl\beta \right) K -\half J \edth\beta 
{}+ \half u \edth \Qp  +\quarter u (\Qp )^2 +\half ur^2G_{mm}.
\label{Gmm:eq}
\end{eqnarray}
These formulae provide the justification for introducing the 
specific forms (\ref{def:HJKQ}) for the NQS connection parameters $(H,J,K,Q)$.

If $\beta$ is known on a null hypersurface $\cN_{z}$ then (\ref{Gll:eq}--\ref{Gmm:eq}) 
form a system of 
differential equations for fields on $S^{2}$ 
along the integral curves of the null 
generators $\ell$ of $\cN_{z}$. 
Although the general ode structure is well known \cite{Sachs62}, 
the form of the RHS is considerably simpler than those found in other
gauges \cite{Bondi62},\cite{NU62},\cite{CMS91}.  Note that the NP equivalents 
of (\ref{Gll:eq}--\ref{Gmm:eq}) are easily determined: for example, 
Eqn.~(\ref{Gll:eq})
is equivalent to the NP propagation identity 
(with $\kappa_{NP}=0$ and $G_{\ell\ell}=2\Phi_{00}$)
\begin{equation}
    D\rho_{NP} = \rho_{NP}^{2} + |\sigma_{NP}|^{2} + 
    2\mbox{Re}(\epsilon_{NP})\rho_{NP} + \Phi_{00},
\label{DrhoNP:eq}
\end{equation}
which is just the null Raychaudhuri equation.

The remaining Einstein equations may be handled via the second Bianchi
identity \cite{Bondi62,Sachs62}.
\begin{Lemma}
Let $\cN$ be an expanding null hypersurface with 
generating null vector $\ell$, and let ${\cal T}$ be a hypersurface which 
intersects each of the null curves of $\cN$ exactly once.
Suppose that $F_{ab}$ is a symmetric 2-tensor satisfying the conservation 
law $F_{ab}^{\ \ ;b}=0$, and 
\[  F_{\ell\ell}=F_{\ell n}=F_{\ell m} = F_{mm}=0,
\]
and such that $F_{nn}=F_{nm}=0$ on ${\cal T}$.
Then $F_{ab}=0$ everywhere on $\cN$.
\end{Lemma}
\Proof    Writing out $F_{ab}^{\ \ ;b}=0$ 
in NP form, using $F_{\ell\ell}=F_{\ell n}=F^{\ell m} = F_{mm} =0$ and the 
bicharacteristic condition $\kappa_{NP}=0$, gives \cite{NP62}
\begin{eqnarray}
    0 & = & \Re\rho_{NP} F_{m\bar{m}},
\label{nc:mmb}
\\
    D_{\ell}(F_{nm}) & = & 
    (2\rho_{NP}+\bar{\rho}_{NP}-2\bar{\varepsilon}_{NP})F_{nm}
    +\sigma_{NP} F_{n\bar{m}}  
\nonumber \\
&&  {}+ D_{\bar{m}}F_{m\bar{m}} + 
    (\bar{\pi}_{NP}-\tau_{NP})F_{m\bar{m}},
\label{nc:nm}
\\
    D_{\ell}(F_{nn}) & = & 
    2\Re(\rho_{NP}-2\varepsilon_{NP})F_{nn}
    +D_{m}F_{n\bar{m}}+D_{\bar{m}}F_{nm}
\nonumber \\
\lefteqn{\hspace*{-2.8em}
  {}-\Re\mu_{NP}F_{m\mbar}
    +\Re((2\beta_{NP}+2\bar{\pi}_{NP}-\tau_{NP})F_{n\mbar}).
}
\label{nc:nn}  
\end{eqnarray}
Since the expansion $\rho_{NP}\ne0$ by hypothesis, Eqn.~(\ref{nc:mmb}) gives 
$F_{m\mbar}=0$; eliminating $F_{m\mbar}$ turns Eqn.~(\ref{nc:nm}) into a 
linear homogeneous ode along the integral curves, which gives $F_{nm}=0$ by 
the initial condition and uniqueness.  Finally, Eqn.~(\ref{nc:nn}) now 
reduces to a linear homogeneous equation for $F_{nn}$ with only the
zero solution 
satisfying the initial condition.  
\QED
When applied to the Einstein tensor, $F_{ab}=G_{ab}$, the lemma shows 
that in order to show that the Einstein equations $G_{ab}=0$ are everywhere 
satisfied, it suffices to find metric parameters satisfying only the 
hypersurface equations (\ref{Gll:eq}--\ref{Gmm:eq}) everywhere, and the 
{\em boundary}
({\em subsidiary} \cite{Sachs62}){\em  equations} $G_{nn}=G_{nm}=0$
on one hypersurface $\cT$ transverse to the expanding null foliation.
Observe that the signature of the hypersurface $\cT$ is immaterial; all 
that is needed is that $\cT$ be transverse to the outgoing foliation.

The following explicit NQS expressions for
$G_{nm}$ and $G_{nn}$ show that the boundary equations may be used to 
constrain the initial data on $\cT$ for the hypersurface equations 
(\ref{Gll:eq}--\ref{Gmm:eq}):
\begin{eqnarray}
r\,\cD_z\left({J/u}\right) &=&
  {v^2}\,r\cD_r\left(J/(uv)\right)
 +\half(\div\gamma - v\,\div\beta)J/u
 \nonumber\\ && {}
  +   2u^{-1}|K|^{2} - 
   \nabla_{Q^{+}}v - \Delta v + u {r^2} G_{nn} 
\label{eq:Gnn}
\\
r\,\cD_z Q^{+} &=&
 {}
(v\,r\cD_r + J -v\edth\bar{\beta}+\edth\bar{\gamma})Q^{+} 
   - K\bar{Q}^{+}
   + 2\edthbar K 
\nonumber\\ && {}
  + 2u^{-1}r\cD_r(u\edth v) 
   -(2 + \imu \curl\beta)\edth v
  {}  +\edth J 
  -2u^{-1}J\edth u
  - 2u r^{2} G_{nm} .
\label{eq:Gnm}
\end{eqnarray}
The terms on the right hand sides of (\ref{eq:Gnn},\ref{eq:Gnm})  are 
determined on a single null hypersurface from the hypersurface equations 
and a gauge choice for $\gamma_{l=1}$, thereby determining the evolution terms
$\partial(J/u)/\partial z$ and $\partial(\Qp)/\partial z$. 
This constrains $(J/u)$ and $\Qp$ on $\cT$ --- clearly when $\cT$ is taken to be 
a level set of $r$ (so $\partial_{z}$ is tangent to $\cT$), whilst for more 
general transverse hypersurfaces $\cT$, using knowledge of 
$\partial(J/u)/\partial r$ and $\partial(\Qp)/\partial r$.

The remaining Einstein tensor equation $G_{m\bar{m}}=0$ is called the 
\emph{trivial} equation in \cite{Sachs62} by virtue of (\ref{nc:mmb}).
In the NQS parameterisation,  $G_{m\bar{m}}$ is given explicitly by 
\begin{eqnarray}
\label{eq:Gmb}
	u r^{2}G_{m\bar{m}} & = & 
	r\cD_{r}J -\half \div \beta\,J - u|Q^{+}|^{2} +\half u \div Q^{+} 
     + \bar{K}\edth\beta + K \edthbar\bar{\beta} +
	         \bar{Q}^{+}\edth u + Q^{+}\edthbar u  
\nonumber 
\\
&  & {}
	 + r^{2}(v \cD_{r} - \cD_{z})(u^{-1}\cD_{r}u)
	        + u^{-1}r^{2}\cD_{r}(u\cD_{r}v).
\end{eqnarray}

\section{Some Applications}

The simplicity of the explicit NQS expressions for the Einstein tensor 
suggests many possible applications, and in the following we outline three: 
the formal structure of the null-timelike boundary value problem 
\cite{Bondi62}; matching conditions across a null hypersurface and the 
propagation of impulsive waves \cite{Stellmacher38}; and the relation 
between the null intrinsic geometry (type I \cite{Penrose72}) and the class 
of affine null geodesic parameterisations.

A formal solution algorithm for the Einstein equations in the NQS gauge 
starts with $\beta$ defined on a null hypersurface $\cN_{z}$.  Specifying 
$H,J,K,\Qm$ on the intersection $\cT\cap\cN_{z}$ with a transverse 
hypersurface gives boundary (initial) conditions for the hypersurface 
equations, with $\beta|_{\cN_{z}}$ providing source terms.  Solving the 
hypersurface equations for $H,J,K,\Qm$ on $\cN_{z}$ and imposing a gauge 
fixing condition on the $l=1$ components of $\gamma$ (eg.~$\gamma_{l=1}=0$) 
then determines the remaining metric parameters $u,v,\gamma$.  The 
definition Eqn.~(\ref{def:HJKQ}) of $\Qm$ determines $\partial\beta/\partial 
z$ on $\cN_{z}$, which is used to evolve $\beta$ to the ``next'' null 
hypersurface $\cN_{z+\Delta z}$.  The boundary equations 
(\ref{eq:Gnn},\ref{eq:Gnm}) determine new boundary data for $J/u$, $\Qp$, which are 
combined with arbitrary boundary data for $u,K$ on $\cT$ and the new seed 
field $\beta|_{\cN_{z+\Delta z}}$ to repeat the process.  Practical 
experience with a direct numerical implementation of this procedure has 
been very good, which suggests that it might be possible to prove existence 
theorems by this approach.

The freedom in the choice of $u$ on $\cT\cap\cN_{z}$ should be interpreted 
as the freedom in choosing the outgoing null hypersurfaces.  This can be 
seen as follows: at the linearised level, the coordinate tangent vector 
$\partial_{z}$ points from one null hypersurface to the ``next'' --- the 
$z$-translations are diffeomorphisms preserving the null foliation, and 
generated by the vector field $\partial_{z}$.  Adding a component 
tangential to the null hypersurfaces does not affect this property.  Since 
the NQS null vector $\ell=\partial_{r}-r^{-1}\beta$ depends only on the 
\emph{intrinsic} (type I) geometry of $\cN_{z}$, the equivalence class of 
spacetime evolution vectors $W$ preserving a given null foliation may be 
characterised by the inner product $g(W,\ell)$.  But for the deformation 
vector $W=\partial_{z}$, this inner product is determined by $u$ since 
$u=-g(\partial_{z},\ell)$.  Hence the choice of $u>0$ on $\cT$ determines 
the intersection 2-surfaces $\cN_{z}\cap\cT$, which in turn generate the 
null hypersurfaces $\cN_{z}$; in other words, the boundary freedom in $u$ 
is a gauge (NQS coordinate) freedom.
 
The boundary data for $K$ 
(essentially the incoming shear $\bar{\lambda}_{NP}$) is unconstrained and 
may be considered as representing the 
outgoing gravitational radiation injected into 
the spacetime from the region beyond the boundary ${\cT}$.
In a similar fashion, we may regard the shear vector $\beta$ (which is 
arbitrarily specifiable on one outgoing surface $\cN_{z}$)
as measuring the incoming 
radiation.

Since $\beta$ acts as a source for the  remaining fields on the null
hypersurface, we might expect NQS metrics with $\beta$ identically zero to 
play a distinguished role.   Such metrics admit a null geodesic 
congruence which is expanding, twist free and shear free, and thus
correspond to the Robinson-Trautman spacetimes \cite{RT62}. 
The NQS parameterisation of Robinson-Trautman metrics with $S^{2}$ 
cross-sections is described in \cite{DIW69} and the remaining NQS gauge 
freedoms are classified in \cite{RAB96b}.


NQS coordinates may also be used to describe the junction conditions for 
null hypersurfaces.  The NQS approach has the advantage of simplicity, when 
compared with general coordinate methods \cite{Stellmacher38,Synge62} 
which must separately consider the effects of coordinate changes on the discontinuity 
fields.  On the other hand, the reliance on a special coordinate system 
apparently limits 
the applicability of the NQS analysis.

Let $M^{\pm}$ be two spacetimes with NQS coordinates and 
having null boundary pieces $\cN^{\pm}$ (one past, the other future), together 
with an identification map $\cN^{-}\simeq\cN^{+}\simeq\cN$.  
We use the notation 
$[\cdot]$ to denote the discontinuity of parameters across the matching 
surface, eg.~$[\beta]=\beta_{\cN^{+}}-\beta_{\cN^{-}}$.  Supposing that the 
matching is an isometry along $\cN$, ie.~$[\beta]=0$ (this is the most natural matching 
criterion), we consider the effect of requiring that the Einstein 
equations be satisfied in the weak (distributional) sense.

Let $\cT$ be a hypersurface transverse to $\cN$ in $M=M^{+}\cup M^{-}$.  
Since we may regard $u|_{\cT}$ as gauge, it is reasonable to require also 
that $[u]=0$ on $\cT\cap\cN$.  Likewise we assume that the $l=1$ spherical 
harmonic components of $\gamma$ are gauged to zero.
The boundary equations (\ref{eq:Gnn}),(\ref{eq:Gnm}) show that $[J/u]=0$, 
$[\Qp]=0$ on $\cT\cap\cN$ since their right hand sides are bounded on 
$\cN$, from which it follows (using the definitions (\ref{def:HJKQ}))
 that $[H]=[J]=[\Qm]=0$ on $\cT\cap\cN$.
Considering the jump in (\ref{Gll:eq}) across $\cN$ shows that $[H]$ 
satisfies the ordinary differential equation
\begin{equation}
r \cD_{r}[H] =
   \left(\half\div\beta -
       \frac{2|\edth\beta|^2+r^{2}G_{\ell\ell}}{2-\div\beta} \right)[H],
\end{equation}
along the null generating curves of $\cN$, and uniqueness with the initial 
data $[H]=0$ on $\cT\cap\cN$ shows that $[H]=0$ on $\cN$.  Since 
$[\div\beta] =0$ we have $[u]=0$ also.
Similar arguments using Eqn.~(\ref{Glm:eq}) show that $[\Qm]=0$ 
(and thus $[Q]=[\Qp]=0$) on $\cN$, 
and using Eqn.~(\ref{Gln:eq}), that $[J]=0$.

However, the discontinuity $[K]$ on $\cT\cap\cN$ 
is not constrained in similar fashion by 
gauge or boundary equation considerations, and thus Eqn.~(\ref{Gmm:eq}) 
yields  a formula for the propagation along $\cN$:
 \begin{equation}
        r\cD_{r}[K] =  (\half\div\beta +\imu\curl\beta)[K].
 \label{shock:K}
 \end{equation}
Using the gauge condition $[\gamma]_{l=1}=0$ on $\cN$ to invert $\cL_{\beta}$, 
from Eqn.~(\ref{Dbeta:def}) we recover the discontinuities across $\cN$ of
the remaining  metric coefficients,
\begin{equation}
\label{[gamma]}
        [\gamma] = \cL_{\beta}^{-1}[K],
\quad
\label{[v]}
        [v]  = -\frac{\div[\gamma]}{2-\div\beta},
\end{equation}
and we find $[\partial\beta/\partial z]\ne0$, using $[Q]=0$ and 
Eqn.~(\ref{def:HJKQ}).  From the  NQS formulas for the Weyl curvature 
spinors
\begin{eqnarray}
\label{Psi0}
    r^{2}u^{-1}\Psi_{0} &=& (r\cD_r + 1 - 2 \edth\bar{\beta})(\edth\beta/u),
\\
\label{Psi1}
    4r^{2}\Psi_{1} & = & (r\cD_r -\edth\bar{\beta})\Qm 
      -\edth\div\beta
\nonumber\\&&{} 
      - (3\bar{Q}+\edthbar u)u^{-1}\edth\beta + 2\edthbar\edth\beta,
\\
\label{Psi2}
    2r^{2}\Psi_{2} &=&  {\textstyle\frac{1}{3}}r^{2}(2G_{\ell n}+G_{m\mbar})
     -(1-\half HJ) 
\nonumber\\&&{} 
     - 2u^{-1}\bar{K}\edth\beta -\half(\edth\Qpbar - 
     \edthbar\Qp),
\\
\label{Psi3}
    4r^{2}u^{2}\bar{\Psi}_{3} & = & uv (r\cD_r -\edth\bar{\beta})\Qp  
    -u(r\cD_z-\edth\bar{\gamma})\Qp 
\nonumber\\&&{} 
      + 2(r\cD_r - 1 - \half\imu\curl\beta)(u\edth v)
\nonumber\\&&{} 
    -u\edth J - (3\bar{Q} -\edthbar u)K - 2u\edthbar K,
\\
\label{Psi4}
    r^{2}u\bar{\Psi}_{4} &=& 
    v^{2}(r\cD_r+1-2\edth\bar{\beta})(K/(uv))
\nonumber\\&&{} 
      - (r\cD_z-2\edth\bar{\gamma})(K/u) + \Qp\edth v +\edth\edth v,
\end{eqnarray}
we see easily that $\Psi_{0},\Psi_{1},\Psi_{2}$ and $\Psi_{3}$ will be 
bounded, but ${\Psi}_{4}$ will have a $\delta$-function component 
$-r^{-1}u^{-2}[\bar{K}]\delta(z)$, giving an impulsive gravitational wave 
\cite{Penrose72} 
propagating along $\cN$.

Although the resulting metric is not continuous in NQS coordinates, there 
is a (non-$C^{1}$) coordinate change which makes the metric continuous, 
since the boundary identification is an isometry \cite{CD87}.  There are 
two interesting points about this NQS construction: it provides a geometric 
coordinate condition which does not detect the ``optimal'' metric 
regularity ($C^{0}$), and secondly, it provides strong evidence for the 
existence of \emph{vacuum} spacetime metrics admitting particular 
coordinates in which the metric is not continuous, but such that the full 
curvature tensor, computed in those coordinates, is well defined as a 
distribution.  Such discontinuities are well beyond the reach of the usual 
Sobolev space approaches to local existence for the Cauchy problem for the 
Einstein equations, for which $g\in H^{2}$ seems to be the best plausibly 
possible (and $g\in H^{5/2+\epsilon}$ is the best proven result at this time).  

The difference between characteristic and $3+1$ 
approaches is again underscored by noting that 
whereas 
the Einstein equations across a null matching
hypersurface may be satisfied with unbounded curvature components, 
determined by the shock transport law (\ref{shock:K}), 
imposing the Einstein equations across a smooth spacelike matching hypersurface  
forces the 
metric to be $C^{2-}$ in Gaussian coordinates \cite{Synge62}.
 
{}From the  dual reformulations of Eqs.~(\ref{Gln:eq}), (\ref{Gmm:eq}), 
\begin{eqnarray*}
r\cD_z (u H) &=&  
 r\cD_r(uvH)   + u r^2 G_{\ell n}  
\\ &&{}   
 -u(1+Hv - HJ)
  + \half u(|\Qm|^{2} - \div \Qm),      
\\
r\cD_z(\edth \beta) &=& 
     r\cD_r (v\edth \beta) - \half  u r^2 G_{mm}
\\ &&{}   
+  \left( - v  + \imu\,\curl \gamma- \imu \,v\,\curl \beta \right) \,\edth \beta 
\\ &&{}   
+  \half(uHK+J\edth\beta)
-  \quarter u (\Qm)^2 + \half u\edth \Qm,
\end{eqnarray*}
which correspond to interchanging the role of the inward and outward null 
directions,
we see that the 
conditions $\div[\beta]=0$, $\edth[\beta]=0$ are required by the Einstein 
equations.  Thus the isometry assumption $[\beta]=0$ amounts to a constraint 
on the remaining, purely rotational, component of $[\beta]$.

Because the NQS foliation is intrinsic to the null hypersurfaces 
(determined by the type I geometry, in the terminology of 
\cite{Penrose72}), 
the interaction between the Einstein tensor and the null geodesic 
parameterisation may be simply explained.  It is easy to see that 
$u^{-1}\ell$ is a null geodesic vector field, and thus the geodesic 
parameterisation is determined by the null type I geometry  (ie.~$\beta$) 
and $u$,  
which is in turn determined from $G_{\ell\ell}=2\Phi_{00}$ by 
Eqn.~(\ref{Gll:eq}), up to a choice of initial condition which amounts to 
fixing the scaling of the affine parameter.  
Thus, the effect of matter (as described by $G_{\ell\ell}\ge0$) is to slow 
down the null geodesic parameterisation, when compared to the parameterisation 
on an isometric null hypersurface in vacuum.  
This leads to a redshift of outgoing photons, as seen by an observer at 
constant NQS radius. 


\end{document}